\documentclass[conference]{IEEEtran}

\usepackage{cite}
\usepackage{amsmath,amssymb, amsfonts, dsfont, braket}
\usepackage{mathtools, braket, lipsum, bibentry}
\usepackage{graphicx, multirow, tabularx, hyperref}
\usepackage{textcomp}
\usepackage[usenames,dvipsnames]{xcolor}
\usepackage[utf8]{inputenc} 
\usepackage[T1]{fontenc}    
\usepackage{booktabs}       
\usepackage{nicefrac}       
\usepackage{microtype}      
\usepackage{float}
\usepackage{enumitem, lipsum}
\usepackage[caption=false,font=footnotesize]{subfig}

\def\BibTeX{{\rm B\kern-.05em{\sc i\kern-.025em b}\kern-.08em
    T\kern-.1667em\lower.7ex\hbox{E}\kern-.125emX}}
\graphicspath{{figures/}}
\newcommand{\washington}{ibmq\_washington}

\hypersetup{colorlinks=true, linkcolor=black, urlcolor=blue}

\graphicspath{{../pdf-tex/figures/}{figures/}}

\begin{document}
\title{Adaptive channel estimation for mitigating circuits executed on noisy quantum devices}

\author{
\IEEEauthorblockN{Samudra~Dasgupta$^{1,2*}$ and Travis S.~Humble$^{1,2\dagger}$}
\IEEEauthorblockA{\textit{$^1$Bredesen Center, University of Tennessee, Knoxville, USA}\\
\textit{$^{2}$Quantum Science Center, Oak Ridge National Laboratory, Oak Ridge, Tennessee, USA}\\
\textit{$^*$sdasgup3@vols.utk.edu, $^\dagger$humblets@ornl.gov}\\
}}

\maketitle

\begin{abstract}
Conventional computers have evolved to device components that demonstrate failure rates of $10^{-17}$ or less, while current quantum computing devices typically exhibit error rates of $10^{-2}$ or greater. This raises concerns about the reliability and reproducibility of the results obtained from quantum computers. The problem is highlighted by experimental observation that today's NISQ devices are inherently unstable. Remote quantum cloud servers typically do not provide users with the ability to calibrate the device themselves. Using inaccurate characterization data for error mitigation can have devastating impact on reproducibility. In this study, we investigate if one can infer the critical channel parameters dynamically from the noisy binary output of the executed quantum circuit and use it to improve program stability. An open question however is how well does this methodology scale. We discuss the efficacy and efficiency of our adaptive algorithm using canonical quantum circuits such as the uniform superposition circuit.
Our metric of performance is the Hellinger distance between the post-stabilization observations and the reference (ideal) distribution.
\end{abstract}




\section{Introduction}
Modern (classical) computers have device components with failure rates of $10^{-17}$ or less. Current state of the art quantum computers however exhibit gate-level (e.g. CNOT) failure rates of around $10^{-2}$ \cite{bravyi2021mitigating}. Consequently, reliability and reproducibility of results obtained from quantum computers is a problem in modern quantum information science \cite{kliesch2021theory},
\cite{coveney2021reliability},
\cite{baker20161},
\cite{nichols2021opinion},
\cite{blume2010optimal},
\cite{ferracin2021experimental}.
\par
The problem is made worse by our experimental observation that today's NISQ \cite{preskill2019quantum} devices are unstable. Moreover, remote quantum cloud servers typically do not provide users with an ability to calibrate the device themselves. Also, the calibration may not be up-to-date (such as due to device drift) 
\cite{proctor2020detecting}, 
\cite{znidaric2004stability}, 
\cite{zhang2021predicting}. 
An additional complication arises due to the fact that the existence of correlations amongst different parts of the circuit may increase error on one part of the circuit while minimizing error elsewhere. Thus, it is often not clear what is the right optimal operating point for the compensated circuit.
\par
Key device calibration parameters often fluctuate with time in-between calibrations (temporal instability) and also across the chip (spatial stability). Such instability cannot be resolved by one-time granular sampling and calibration. The quantum noise channel in fact is a random variable that needs adaptive treatment. Using inaccurate characterization data for mitigation can have devastating impact on circuit reproducibility.
\par
Using the most recent device calibrations and hoping this will be accurate in-the-mean, albeit with large error bars, can be erroneous 
\cite{proctor2020detecting}. 
Instead of using the last available noise channel characterization data, one might think that a better approach is higher-frequency characterizations and using the most recent one and hoping the results are more stable and accurate. But this just passes on the problem to a different time-scale. Moreover, characterizing dozens of model parameters throughout the day for various qubit combinations is resource intensive and may not be feasible.
\par
As we shall see, a sounder approach is to dynamically infer the shift in relevant channel parameters using Bayesian inference \cite{lukens2020practical},
\cite{zheng2020bayesian},
\cite{gordon1993novel},
\cite{kotecha2003gaussian}.
We ask whether one can infer the channel parameters dynamically from the noisy binary output of the quantum circuit execution and use it to mitigate the errors (using a closed feedback loop). Preliminary results suggest that we might be able to make the experiments conducted on unstable devices more reproducible using this adaptive scheme even without the knowledge of characterization data. An open question however is how well does this methodology scale. We seek to discuss the efficacy and efficiency of our adaptive algorithm using a few canonical quantum circuits such as the uniform superposition creator. 
Our metric of performance is the Hellinger distance between the post-stabilization observations and the reference (ideal) distribution.

\section{Theory}
\subsection{Ideal quantum computer}
An idealized quantum computer \cite{nielsen2002quantum} comprises an $n$-qubit register that encodes a $2^{n}$-dimensional Hilbert space $\mathbb{C}^{2\otimes n}$ and a set of operations that transform the register. Physical realization of such an idealized information processor however introduce noise processes that must be accounted for in the operational description \cite{kliesch2021theory, ferracin2021experimental, coveney2021reliability, blume2010optimal}. Such noise manifest as errors in the computational output, and methods have been developed to mitigate against such errors. This includes fault-tolerant operations based on quantum error correction techniques as well as mitigation methods based on pre- and post-processing of the computational output.
\subsection{Noisy real device}
An operational description for noisy quantum circuits can be provided using quantum channel formalism:
\begin{equation}
\rho' = \mathcal{E}_e(\rho)= \sum\limits_{k} M_k \rho M_k^\dagger,
\end{equation}
where $\mathcal{E}_e(\cdot)$ is a channel operator expressed in terms of Kraus operators $\{M_k\}$, $e$ is the error parameter characterizing the noise channel and $\rho$ represents the state of a quantum register in the absence of noise.
\subsection{Experimental evidence for channel variability}
\begin{figure*}
  \centering
  \begin{tabular}{ c @{\hspace{40pt}} c }
      \includegraphics[width=.8\columnwidth]{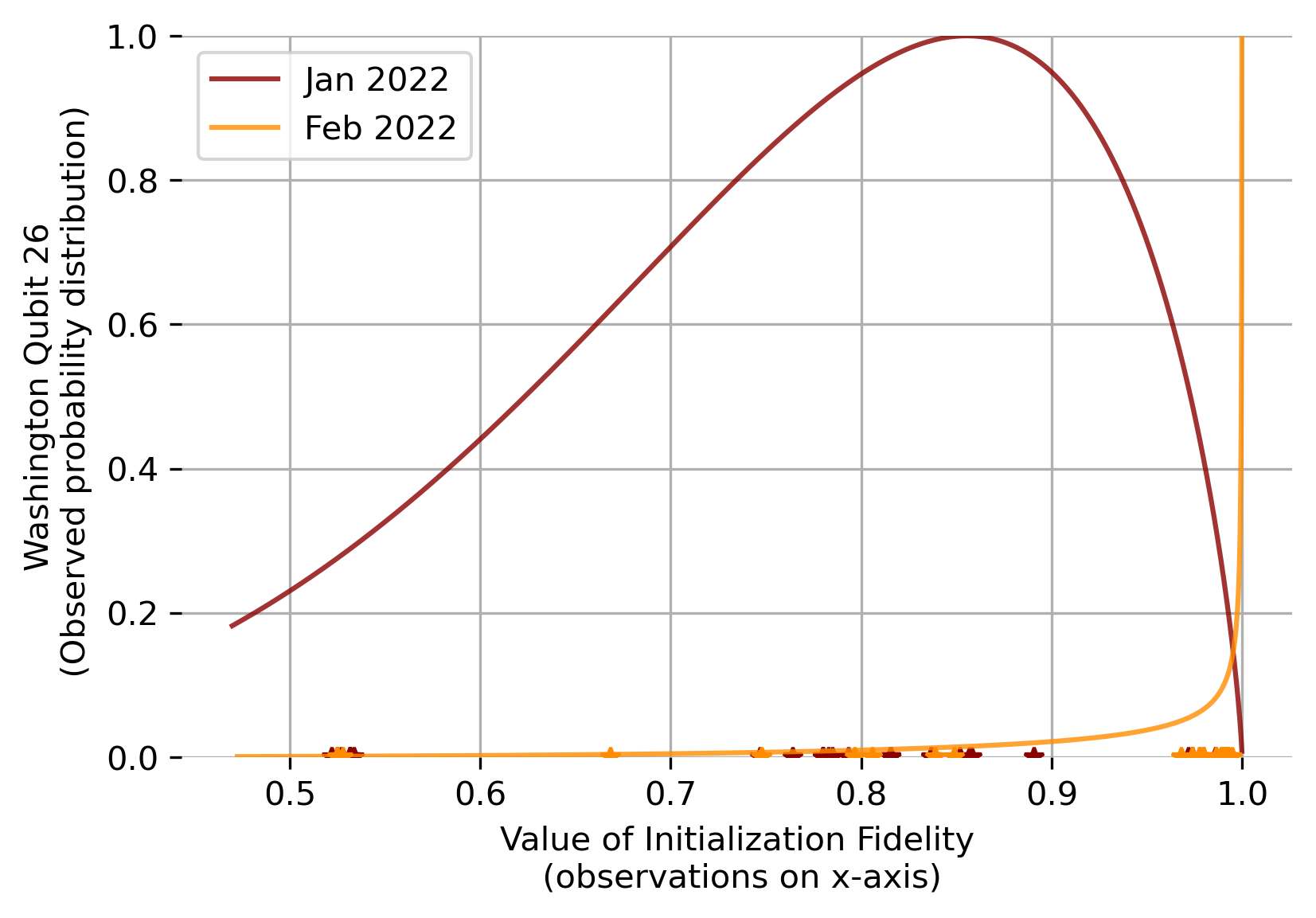} &
    \includegraphics[width=.95\columnwidth]{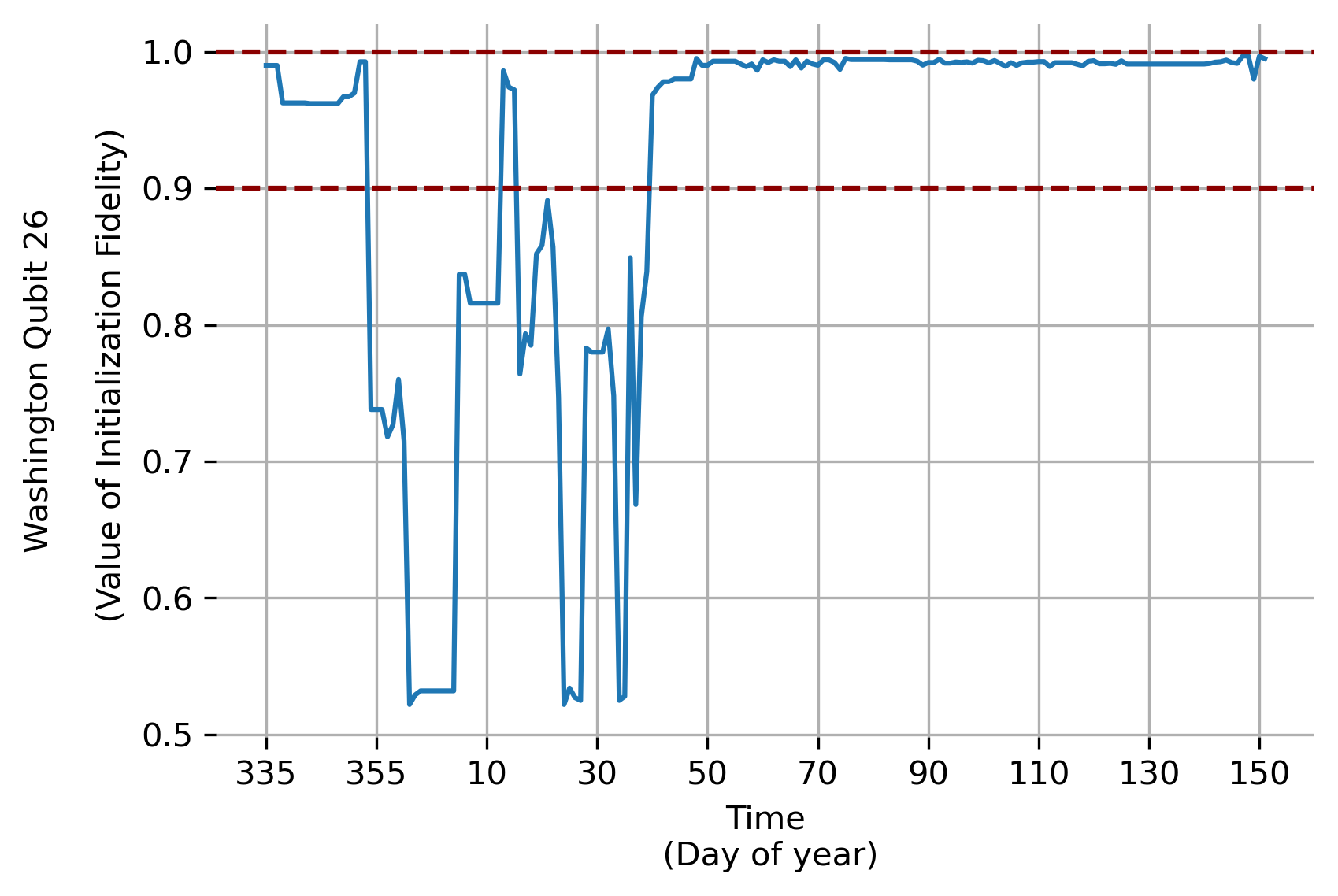} \\
    \small (a) &
      \small (b)
  \end{tabular}
  \medskip
  \caption{ (a) Wide variance seen in the distribution of Initialization Fidelity, indicating an unstable readout channel. Data shown for register element \#26. (b) Dynamic variation for the same register element for the period Dec 2021-May 2022.
}
\label{fig:q26_washington_dec_may2022}
\end{figure*}

The assumption that quantum noise channels in today's NISQ (noisy intermediate scale quantum) devices \cite{preskill2019quantum} can be characterized by constant error parameters 
does not hold up to experimental scrutiny. For example, consider the single-qubit readout error channel for register element \#26 in the \washington ~device as shown in Fig.~\ref{fig:q26_washington_dec_may2022}. Here, the fidelity metric observed during two different periods of time has been fit to a beta distribution, indicating stark 
variability of the readout channel behavior during these two execution windows. Not only is the mean error parameter $e$ fluctuating with time, even the distribution of the random variable $e$ keeps changing.
%
\subsection{Quantifying stability of a quantum program}
In light of the experimental evidence for channel variability, 
we quantify the stability of a quantum program with respect to fluctuations in the parameters of a quantum channel as follows:
\begin{equation}
\mathds{E}_{e} \left[ d_{\mathcal{H}}^{e}  \right] < \epsilon, \hspace{4mm} 
\forall \mathcal{C}, t
\label{eq:stability_stringent}
\end{equation}
where $d_{\mathcal{H}}^{e}$ is the Hellinger distance between the experimentally observed and the reference output distribution, $\mathcal{C}$ is any quantum circuit, $t$ denotes time and  
\begin{equation}
\mathds{E}_e (\cdot) = \int\limits_e (\cdot) p(e) de
\end{equation}
is the expectation operator with respect to the distribution of error parameter $e$. The Hellinger distance $d_{\mathcal{H}}$ between $f$ and the reference distribution $g$ is defined as $d_{\mathcal{H}}(f, g) = \sqrt{1-\textrm{BC}(f, g)}$ where the Bhattacharyya coefficient is given by  $\textrm{BC}(f, g) = \sum_{i}{\sqrt{f_i g_i}}$ with $f_{i}, g_{i}$ the $i$-th discrete output of these distributions. It is notable that the Hellinger distance vanishes when the distributions are identical and grows to unity as the distributions become completely disjoint.
\par 
For an observable $O$, the stability condition can be stated as
\begin{align}
\mathds{E}_e\left[\{ \braket{O^{noisy}_e} - \mathds{E}_e[ \braket{O^{noisy}_e}] \}^2\right] <& \epsilon, \hspace{4mm} \forall \mathcal{C}, t
\end{align}
where
\begin{equation}
\mathds{E}_e[ \braket{O^{noisy}_e} ] = 
\int\limits_e \sum\limits_m m \textrm{Tr}[\mathcal{E}_{e}(\mathcal{C}\rho \mathcal{C}^\dagger)\Pi_m]p(e) de
\end{equation}
\subsection{Adaptive scheme for stabilization}
We propose an adaptive stabilization scheme that enables unstable devices to be used for executing quantum circuits with stable results. The basic algorithm learns and updates the channel parameters using Bayesian inference, given only the digital output from the quantum computer and uses the information for adaptive error compensation This helps to reduce the output fluctuation attributable to time-varying noise.
\par 
Let $y_k = (s_1^k\cdots s_n^k)$ denote the $n$-bit digital output of a quantum computer measured in the computational basis at the $k$-th circuit execution with $s_i^k \in \{0,1\}$ being a classical bit. Let $y_{1:S}(t)$ denote the $S$ samples collected from the $S$ ensemble executions (sometimes called \textit{$n$-shots}).
\par 
As before, $e$ is the error parameter vector characterizing the quantum error channels, which are linear maps in the $2^n \times 2^n$ expanded Hilbert space (also called the Liouville space). For example, for a depolarizing channel for a single qubit register, $e\in [0, 1]$ 
and its effect is given by the super-operator $\mathcal{E}_e(\cdot)$ with:
\begin{equation}
\mathcal{E}_e(\rho) = (1-e)\rho + \frac{e}{3}X\rho X + \frac{e}{3}Y\rho Y + \frac{e}{3}Z\rho Z
\end{equation}
where $X,Y,Z$ are the Pauli matrices. Let $p(e; t$) be the joint probability distribution function for the error parameter $e$ at time $t$. Then from Bayes theorem:
\begin{align}
\textrm{Pr}[e; t| y_{1:S}(t)] &\propto \textrm{Pr}[y_{1:S}(t) | e] p(e; t)
\end{align}
where, 
\begin{itemize}
\item $\textrm{Pr}[y_{1:S}(t) | e]$ is the likelihood whose computation requires a noise model assumption

\item $p(e; t)$ is the prior probability distribution which can be estimated from the device characterization data available (from time $0$ to time $T \leq t-1$) as the starting point. Alternately, the prior can be a simple uniform distribution to signify the complete lack of information about the quantum channels. 

\item $\textrm{Pr}[e; t| y_{1:S}(t)]$ is the joint posterior distribution for the error channel  given the $S$ observations at time $t$; and,

\item the proportionality constant $c$ is given by:
\begin{align}
\frac{1}{c} =&  \int\limits_e p(y_{1:S}(t) | e) p(e; t)de
\end{align}
It may be possible to calculate $c$ analytically under rare circumstances. Generally, it is computationally intractable. However, it is not required when using a Markov Chain Monte Carlo (MCMC) method such as Metropolis-Hastings \cite{robert1999monte, hastings1970monte, shang2015monte}.

\end{itemize}
The next step in the algorithm is to obtain the maximum-a-posteriori (MAP) estimate for $e$:
\begin{align}
\hat{e} =&  \underset{e}{\mathrm{ argmax \hspace{2mm} }} p[e|y_{1:S}(t)]
\end{align}
This serves as the informationally complete latest update which can be used to correct biases in gate operations from the software interface. The ability to scale this approach is determined by the granularity of the channel parametrization. Using a large number of parameters that exponentially scale with the register elements may not be necessarily useful for statistical estimation purposes (and often degrades the results). Simple models, which scale favorably, often lower bias and avoid over-fitting \cite{vapnik1999nature}.
\section{Application}
For example, consider a 4-qubit register initialized to $\ket{0000}$. Each  register element is subsequently subjected to a Hadamard gate. The channel operator acting in the Liouville space is characterized by the error parameter $e \in \mathds{R}^{12}$ which has twelve elements. This is because each register element is affected by its own set of three unique parameters: $e_{0,i} =$ readout error when the channel input state is $\ket{0}$ for register element $i$, $e_{1,i} =$ readout error when the channel input state is $\ket{1}$ for register element $i$ and $e_{2,i} =$ the Hadamard gate error (in radians) for register element $i$. The ingredients of the Bayesian model specification are then as follows. 
Assuming there is no cross talk between the register elements and that the three sources of error described are independent, then to a first-order approximation, the prior is given by:
\begin{align}
p(e^{\textrm{circuit}})&=p(e_{\textrm{(channel 1)}})
p(e_{\textrm{(channel 2)}})\\ \nonumber
&p(e_{\textrm{(channel 3)}})
p(e_{\textrm{(channel 4)}})
\end{align}
where,
\begin{align}
p(e_{\textrm{(channel i)}}) &= p(e_{0,i}, e_{1,i}, e_{2,i})\nonumber\\
&= \prod\limits_{k=0}^{2} p(e_{k,i}) \textrm{    (assuming independence)}
\end{align}
and,
\begin{align}
p(e_{k,i}) &= p(e_{k,i}; \alpha_{k,i}, \beta_{k,i}) \nonumber \\
&= \frac{e_{k,i}^{\alpha_{k,i}-1} (1-e_{k,i})^{\beta_{k,i}-1}}{\textrm{Beta}(\alpha_{k,i}, \beta_{k,i})}
\end{align}
where $\textrm{Beta}(\cdot,\cdot)$ is the beta function:
\begin{equation}
\textrm{Beta}(x,y)=\int\limits_0^1 t^{x-1}(1-t)^{y-1}, \hspace{4mm} \forall x,y>0
\end{equation}
The likelihood is obtained as:
\begin{align}
\textrm{Pr}[y_{1:S}(t) | e] &= \prod \limits_{k=1}^{S} \textrm{Pr} (y_k | e)\\
&= \prod\limits_{k=1}^{S} \prod\limits_{j=1}^{n} \pi_{0,j}^{1-s_j^k}(1-\pi_{0,j})^{s_j^k}
\end{align}
where,
\begin{align}
\pi_{0,j} &= \frac{1+e_{1,j}-e_{0,j}}{2} - \sin(2e_{2,j}) \frac{e_{0,j}+e_{1,j}-1}{2}
\end{align}
The posterior, given below, is estimated using Metropolis-Hastings (as the analytical approach is intractable).
\begin{align}
\textrm{Pr}[e; t| y_{1:S}(t)] &\propto 
\prod\limits_{j=1}^{n} \left[ \pi_{0,j}^{S-\sum\limits_k s_j^k}(1-\pi_0)^{\sum\limits_k s_j^k}\nonumber \right.\\
&\left. \times
\prod\limits_{m=0}^{2}
\frac{e_{m,j}^{\alpha_{m,j}-1} (1-e_{m,j})^{\beta_{m,j}-1}}{\textrm{Beta}(\alpha_{m,j}, \beta_{m,j})}
\right]
\end{align}
Next, the MAP estimate is obtained using a log maximization:
\begin{align}
\hat{e} &= \underset{e}{\mathrm{argmax }} \log \textrm{Pr}[e; t| y_{1:S}(t)] 
\end{align}
Finally, the estimates of $\hat{e}$ are used for a more accurate readout error mitigation (using matrix inversion) and Hadamard compensation for calibration error (using $\pi/4-e_2$ as the angle input instead of $\pi/4$). 
\subsection{Implementation details}
\begin{figure*}
  \centering
  \begin{tabular}{ c @{\hspace{80pt}} c }
      \includegraphics[width=.64\columnwidth]{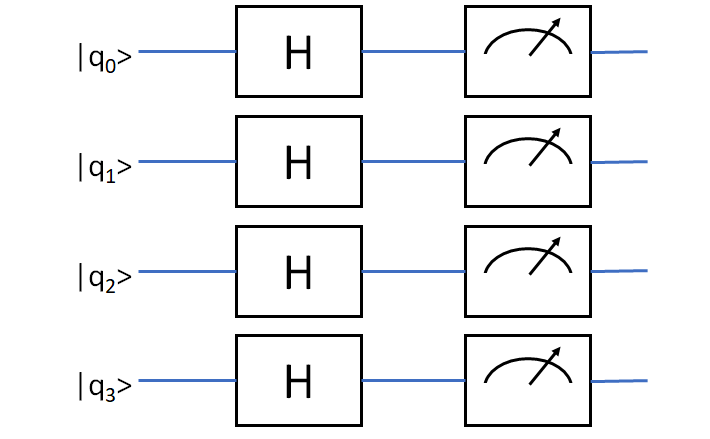} &
    \includegraphics[width=.64\columnwidth]{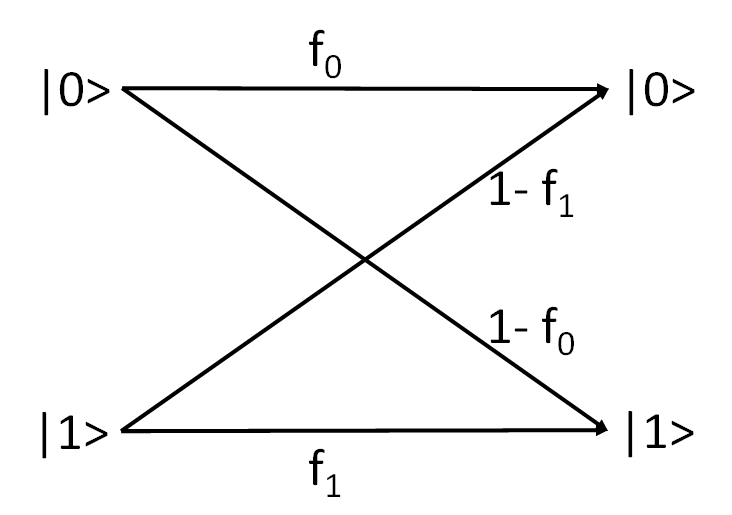} \\
    \small (a) &
      \small (b)
  \end{tabular}
  \medskip
  \caption{ (a) The 4-qubit quantum circuit used for the simulation experiment. Each qubit is assumed to have a different, independent 
  readout error process. Each Hadamard gate is similarly assumed to have a different, independent 
  gate error process. (b) A binary asymmetric readout channel is parametrized by two terms: $f_0$ captures the fidelity of observing $\ket{0}$ when the initial state is $\ket{0}$ and $f_1$ captures the fidelity of observing $\ket{1}$ when the initial state is $\ket{1}$. 
}
\label{fig:setup}
\end{figure*}
We simulated our example above using the Qiskit \cite{ibm_quantum_experience_website} software. The circuit is shown in Fig.~\ref{fig:setup}~(a). The readout error channel was simulated using a binary asymmetric channel (Fig.~\ref{fig:setup}~(b)) for each register element. The error probabilities fluctuate with execution instance. 
The values are drawn from a beta distribution with mean and variance specified as input. The values used for the mean of $e_0$ were $(0.9,  0.8,  0.85, 0.75)$ for each of the register elements respectively. The standard deviation was simply one-tenth of the mean. The values used for the mean of $e_1$ were $(0.85, 0.75, 0.80, 0.70)$ for each of the register elements respectively. Like before, the standard deviation of $e_1$ was simply one-tenth of the mean.
\par 
The 
gate error was simulated using a custom Hadamard gate built using the $u3$ gate which takes as inputs three parameters $\theta, \phi$ and $\lambda$. We set $\theta = \pi/2 + e_2, \phi=0$ and $\lambda=\pi$. Like before, $e_2$ is the 
error term which fluctuates with execution instance. It is assumed to be drawn from a beta distribution with mean (in degrees) = $(3.1,  4.1,  4.9,  2.9)$ for each of the register elements respectively. Also, as before, the standard deviation of $e_2$ was assumed to be simply one-tenth of the mean in each case.
\par 
Each execution assumed an ensemble size of $8192$ (also called $n$-shots). The circuit was executed 10 times for each of the three cases: unmitigated, static mitigation and adaptive mitigation. We describe these three cases next.
\par 
When calculating the Hellinger distance of the output for the \textit{unmitigated} case, we do not do any error compensation or post-processing. The circuit is simply executed and the resulting histogram compared with ideal. Note that during each circuit execution, the noise is assumed to remain constant.
We do 10 runs and the noise fluctuates 
across these runs as per the scheme discussed before. 
\par 
When calculating the Hellinger distance of the output from the \textit{static}  mitigation case, we proceed as follows: 
we assume we have done an excellent job in device characterization and hence know the true mean of the error distribution. We run the circuit with a compensated $u3$ gate where instead of $\pi/2$, we input $(\pi/2 - e_2^{\textrm{true mean}})$ in the script as the input. This serves the purpose of static coherent compensation. Finally, using $e_0^{\textrm{true mean}}$ and $e_1^{\textrm{true mean}}$ we perform readout error mitigation (using simple matrix inversion).
\par 
When calculating the Hellinger distance of the output from the \textit{adaptive}  mitigation case, we proceed as follows: First we run the circuit and observe the binary output after each execution (which has 8192 shots). We do not assume any prior knowledge of the noise characteristics. We instead estimate the empirical distributions of the errors from the output using the framework discussed before. Then we find the MAP estimate for $\hat{e}$ (12 parameters) after each execution. After that, we rerun the circuit with a compensated $u3$ gate where instead of $\pi/2$, we input $\pi/2 - \hat{e_2}$ as the input for each fresh execution. This serves the purpose of adaptive coherent compensation as the noise parameter-mean estimate keeps getting updated after each circuit execution. Finally, using $\hat{e_0}$ and $\hat{e_1}$ we perform readout error mitigation (using simple matrix inversion) on the results from the compensated circuit.
\par
\begin{figure*}
  \centering
  \begin{tabular}{ c @{\hspace{40pt}} c }
      \includegraphics[width=.8\columnwidth]{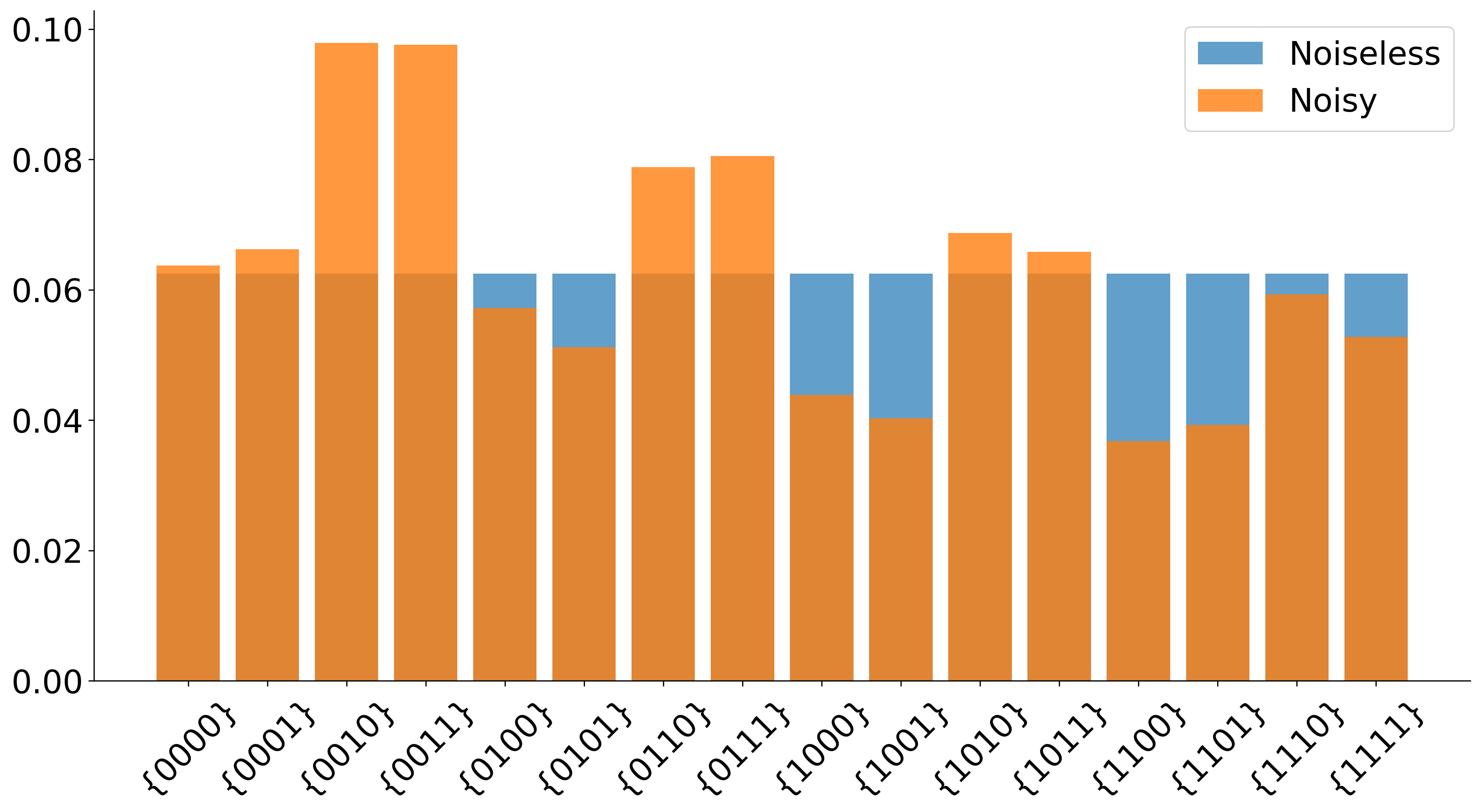} &
    \includegraphics[width=.95\columnwidth]{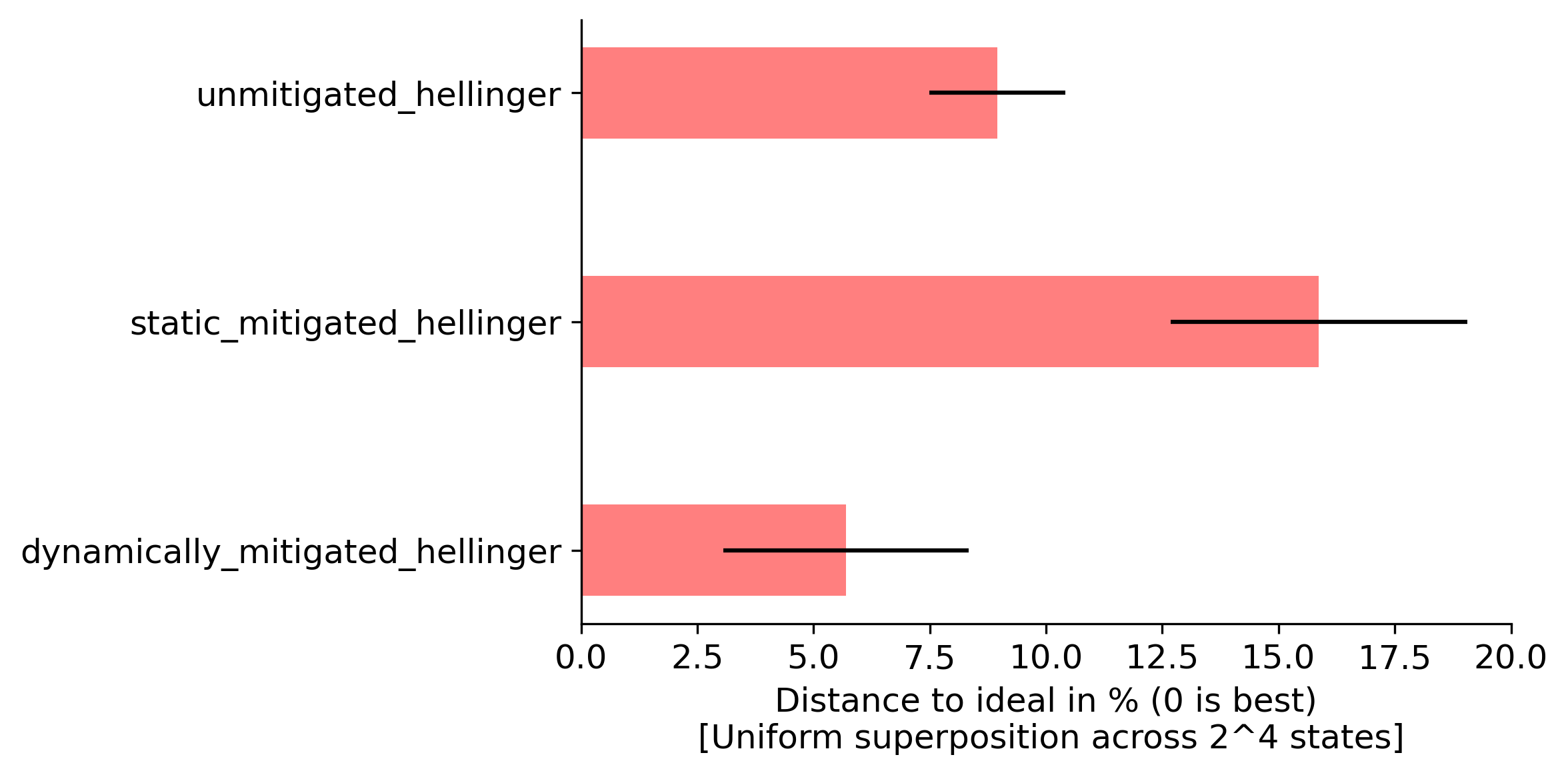} \\
    \small (a) &
      \small (b)
  \end{tabular}
  \medskip
  \caption{ (a) We simulated 
  readout and gate error channels for a quantum circuit with 4 qubits that creates a uniform superposition across all the computational basis states using Hadamard gates. 
The blue bars represent the probability distribution across the measurement outcomes for the ideal, noiseless circuit while the orange bars represent the same for a realization of an execution on an unstable device. (b) The plot shows the distance reduction achieved post adaptive stabilization in presence of 
time-varying noise. The static mitigation worsens the results in presence of time-varying 
noise. The number of samples used in the estimation procedure is held constant at $10^{4}$.
}
\label{fig:q_stabilization}
\end{figure*}

 
Fig.~\ref{fig:q_stabilization} shows a reduction in the circuit error (as measured by Hellinger distance) when an adaptive approach is deployed.
%
\section{Conclusion}
Noise characterizations are assumed to be constant in current research in quantum computing when applying error mitigation. Thus, a natural question arises as to how sensitive is the output to  
variable noise processes. In this study, we have used an adaptive scheme using Bayesian inference to dynamically estimate the noise channels, and used that knowledge to compensate the error. This helps stabilize the circuit in terms of minimizing fluctuations with respect to the time-varying noise. The algorithm does not require a-priori knowledge of channel calibration. In fact, erroneous channel calibration knowledge is acceptable too - as the method learns dynamically from the measurements (binary strings) observed. Our work fills a gap for utilizing unstable quantum platforms with the goal of improving reproducibility in the burgeoning field of quantum information science.

\section*{ACKNOWLEDGMENTS}
\small{This work is supported by the U.~S.~Department of Energy (DOE), Office of Science, National Quantum Information Science Research Centers, Quantum Science Center and the Advanced Scientific Computing Research, Advanced Research for Quantum Computing program. This research used computing resources of the Oak Ridge Leadership Computing Facility, which is a DOE Office of Science User Facility supported under Contract DE-AC05-00OR22725. The manuscript is authored by UT-Battelle, LLC under Contract No.~DE-AC05-00OR22725 with the U.S. Department of Energy. The U.S.~Government retains for itself, and others acting on its behalf, a paid-up nonexclusive, irrevocable worldwide license in said article to reproduce, prepare derivative works, distribute copies to the public, and perform publicly and display publicly, by or on behalf of the Government.  
The Department of Energy will provide public access to these results of federally sponsored research in accordance with the DOE Public Access Plan. 
http://energy.gov/downloads/doe-public-access-plan.}

\bibliographystyle{unsrt}
\bibliography{my_references.bib}

\end{document}